# Insights into the magnetocaloric effect of Gadolinium: A DFT Exploration of Structural, Electronic, and Magnetic Features in Bulk and Film configurations

A. Endichi[1], H. Bouhani[1,2*], O.Baggari[1], H. Zaari[1], O.Mounkachi[1], A. El Kenz[1], A.Benyoussef[3]

[1]Laboratoire de Matière Condensée et Sciences Interdisciplinaires (LaMCScI), Facultés des Sciences P.B. 1014, Université Mohammed V in Rabat, Morocco.

[2]Centrale Casablanca, Centre de Recherche Systèmes Complexes et Interaction, Bouskoura 27182, Morocco

[3] Hassan II Academy of Science and Technology, Rabat, Morocco

*Corresponding author: hamza.bouhani@centrale-casablanca.ma

## Abstract

Gadolinium stand as the favored choice among magnetic refrigerant materials for numerous active magnetic regenerator (AMR) prototypes due to its remarkable ability to exhibit a substantial change in magnetic entropy. This unique characteristic arises from its status as one of the elemental ferromagnets with a high Curie temperature, closely aligning with room temperature conditions, and undergoing a second-order magnetic phase transition. In this comprehensive study, we employ density functional theory (DFT) calculations to explore the structural, electronic, and magnetic properties of both Gadolinium bulk and film configurations. Our primary objective is to gain a deeper understanding of the intricate physics underlying the intriguing magnetocaloric features observed in Gadolinium. This investigation provides valuable insights into the potential applications and the broader implications of Gadolinium in the realm of magnetic refrigeration technology.

## I. Introduction

The conventional vapor compression refrigeration method necessitates the utilization of environmentally harmful refrigerants. Magnetic refrigeration presents itself as a potentially more cost-effective, environmentally friendly, and efficient alternative to traditional refrigeration [1,2]. However, the implementation of this technology requires materials possessing specific properties. The exploration of novel materials for magnetic refrigeration is an actively pursued research area as an example of these research: The chapter book [3] delves into an extensive magnetocaloric materials library, encompassing diverse categories such as lanthanide metals, binary lanthanide-metalloid compounds and high entropy alloys [4]. It introduces a directed search strategy for designing intermetallics that balance criticality and properties, with a specific emphasis on magnetocaloric (MC) effects in magnetic solids for magnetic refrigeration (MR) technology. Notably, Gd2MgTiO6 [5] is highlighted as a noteworthy example, exhibiting a B-site ordered monoclinic double perovskite crystal structure and displaying significant reversible cryogenic MC effects. Furthermore, the book explores RE-TM-X materials, exemplified by RECo12B6 compounds [6], to investigate their intriguing magnetic characteristics. These materials demonstrate a ferrimagnetic ground state, originating from the spin-polarized 3d electrons of Co atoms and influenced by the 4f electrons of RE atoms. This comprehensive examination contributes to a deeper understanding of the magnetic properties of these compounds, furthering the development of materials with potential applications in advanced technologies such as magnetic refrigeration.

The advancements in numerical calculation methods for electronic structure now enable the theoretical study of materials before their synthesis in the laboratory [7]. In the pursuit of enhancing magnetic refrigerator performance through the investigation of new materials, research efforts have also been directed towards the development of innovative refrigerator models, like $Mn_{30}Fe_{20-x}Cu_xAl_{50}$ alloys which a rare-earth-free magnetocaloric materials derived from MnAl-based permanent magnets, exhibit a stable crystal structure, tunable magnetic phase transition temperature, and promising magnetocaloric performance, offering a novel and practical option for room-temperature magnetic refrigeration applications [8].

Various refrigeration cycles have been devised for applications around room temperature, with pure gadolinium established as the experimental reference element [9,10]. At its Curie temperature of approximately 294 K, pure gadolinium exhibits magnetocaloric properties (variation of magnetic entropy $\Delta S_m$ and adiabatic temperature change $\Delta T_{ad}$) of approximately -10 J.kg-1.K-1 and 12 K, respectively, under a magnetic field variation from 0 to 5T [11-12].

The pronounced magnetocaloric effect is attributed to its high magnetic moment (7 $\mu_B$/atom) derived from its 4f[7] electrons. Additionally, the absence of spin-orbit coupling (L=0) in gadolinium eliminates the hysteresis in magnetization/demagnetization cycles [13-14,18]. Conversely, the magnetocaloric effect (MCE) on magnetic thin films is of particular interest for micro-refrigeration devices [15]. In this specific aspect, the intrinsic characteristics of thin films are worthy of investigation. This article focuses on the examination of Gadolinium in bulk and thin film form using density functional theory (DFT) to assess their suitability for magnetic refrigeration.

## II. Theoretical model and Computational details

We conducted first-principles calculations within the Generalized Gradient approximation (GGA), utilizing the full potential linearized augmented plane wave (FP-LAPW) method as implemented in the Wien2k code. Brillouin zone integrations were carried out using 400 k-points. Converged solutions were obtained with RMT*Kmax = 8, where RMT represents atomic sphere radii, and Kmax is the maximum value of the wave vector (K = k + G). The self-consistent calculations are considered to converge when the total energy of the system stabilizes within $10^{-5}$ Ry.

## II. Results and discussions

### 1. Crystal structure and density of state

The magnetic structure Gadolinium (Gd), with electron configuration $4d^{10} 4f^7\ 5s^2\ 5p^6\ 5d^1\ 6s^2$, exhibits a hexagonal structure within space group P63/mmc settings. The standardized atom coordinates are Gd1: 1/3, 2/3, 1/4, and Gd2: 2/3, 1/3, ¾ (refer to Figure 1a and Figure 1b). The unit cell possesses lattice parameters a = b = 3.639 Å and c = 5.812 Å.

To elucidate the nature of the Gadolinium system and the distribution of electrons in the valence and conduction bands, we conducted calculations for the partial and total density of states (DOS) of the Gadolinium system, as illustrated in Figure 2.

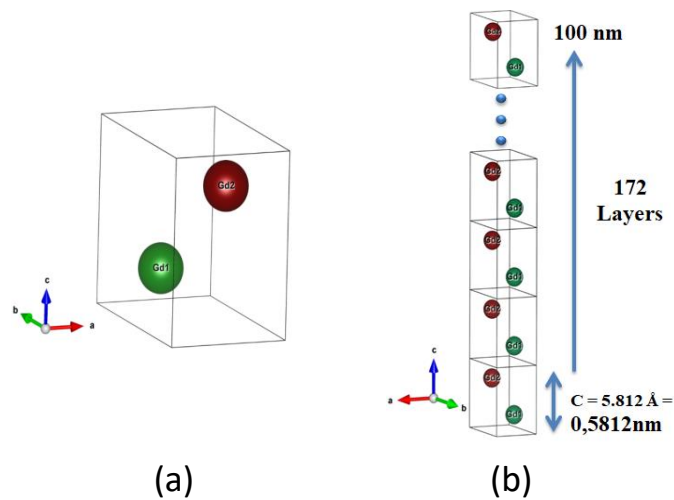

(a)          (b)

**Figure 1.** Hexagonal structure of the Gadolinium (Bulk) and b) Simplified diagram of the layer.

To elucidate the impact of electronic structure on magnetic properties, we calculated the total and partial densities of states (PDOS) for Gadolinium Bulk and a Thin Film (2nm) using the GGA+U approximation, as depicted in Figure 2(d) and Figure 2(e) respectively. The lower valence bands, spanning from (-5.0 eV) to (-4.0 eV), arise from the Gd-f states. The predominant contribution to magnetism originates from the 4f states of Gd atoms. The density of states illustrates the distribution of electrons in the conduction and valence bands of gadolinium.

It is evident that electrons in the valence band occupy localized states, making them unable to participate in electrical conduction phenomena. Conversely, electrons in the conduction band, particularly the free electrons of 4f, are delocalized and actively contribute to electronic conduction. The magnetic moment values for Gd Bulk and Gd Thin Film are 6.863 $\mu_B$ and 7.17641 $\mu_B$, respectively.

The conduction band is partially occupied, which allows the electrons of this band to pass to higher energy levels, without violating the Pauli Exclusion Principle, and thus participate in the conduction.

## 2. Coupling calculation between Gadolinium atoms

The magnetic coupling (J) between two electrons situated on the magnetic orbitals is determined using ab initio methods. Through these calculations, we compute the value of the magnetic coupling ($J_i$), which is then implemented in our system's Hamiltonian as follows:

$$H = -J_1 \sum_{\langle ij \rangle_{1mp}} S_i S_j - J_2 \sum_{\langle ij \rangle_{2Ep}} S_i S_j - J_3 \sum_{\langle ij \rangle_{3Ep}} S_i S_j - \Delta \sum_i S_i^2 - H \sum_i S_i$$

Based on the crystal structure of this system, see Figure 2 (a),(b,) and (c,) we consider 3 types of magnetic coupling in this theoretical study, namely:

$J_1$: The interaction between the first nearest neighbors of Gd1 atoms in Plan 1.

$J_2$: The interaction between second nearest neighbors of Gd2 atoms in Plan 2.

$J_3$: The interaction between Gd1 (Or Gd2) atoms in different plan.

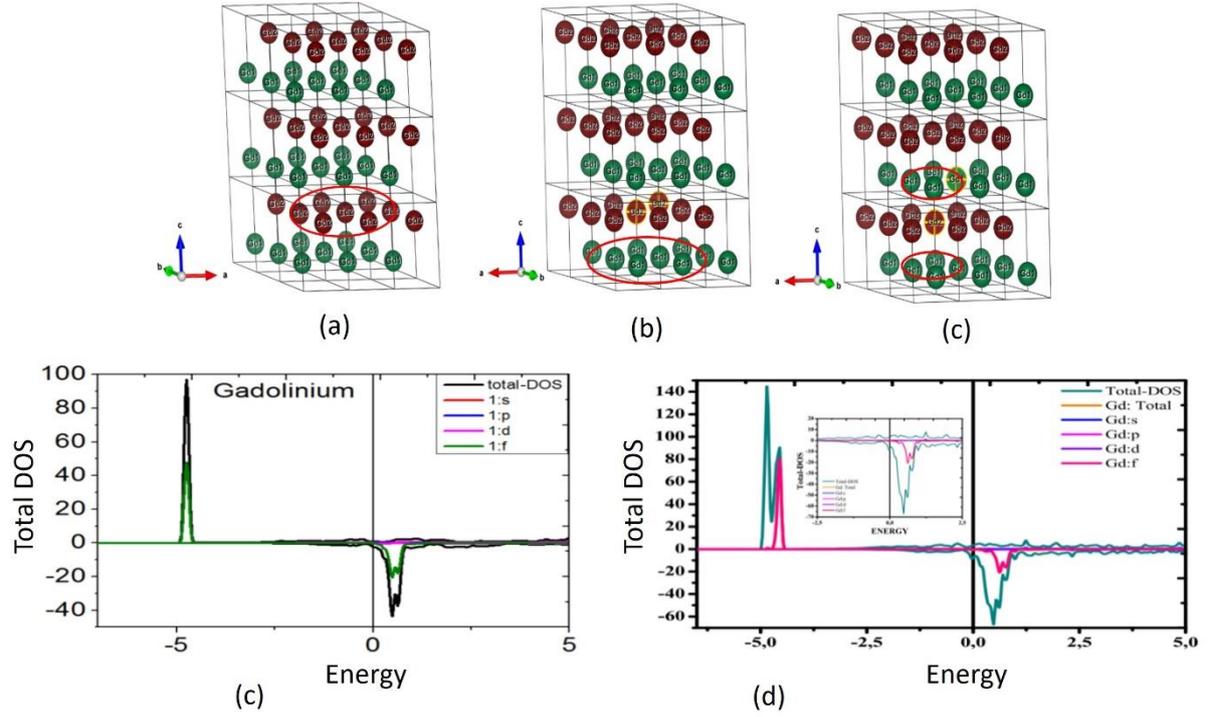

**Figure 2**. a) The coupling between Gd1 atoms, b) The coupling between Gd Atoms and c) the coupling between Gd1 and Gd2 atoms. Total and partial density of states for Gadolinium Bulk (d) and thin film (e) with the GGA+U approximation.

|  | Plan 1($Gd_1$) | Plan 2($Gd_2$) | Between $Gd_1$ and $Gd_2$ |
|---|---|---|---|
| **Distance d(Å)** | 3.6336 | 3.6336 | 3.5715 |
| **Near neighbors z** | 6 | 6 | 6 |
| **Coupling J(eV)** | $1{,}6137.10^{-3}$ | $1{,}6137.10^{-3}$ | $1{,}5728.10^{-3}$ |

**Table 1**. The values of the distance, the near neighbors and the couplings between the atoms of Gd.

The values of magnetic coupling presented in Table 1 are computed using the following relation:

$$J = \frac{1}{2zS^2} \Delta E_{FM-AFM}$$

## 3. Magnetic anisotropy calculation: Easy and hard axis of magnetization

To comprehend the anisotropy exhibited by the Gadolinium compound, we employed the DIPAN package within the WIEN2k program to calculate the system's energy along each axis.

This calculation involves determining the total energy for various axes to identify the axis with the minimal energy. For Gadolinium Bulk, the lowest energy corresponds to the c-axis.

| Axes | Gd Bulk $E_{an}(J/m^3)$ | Gd 0.5nm $E_{an}(J/m^3)$ | Gd 2nm $E_{an}(J/m^3)$ |
|---|---|---|---|
| 001 | -0,72999 | 0,16585 | 0,8578417 |
| 100 | 0,364998 | -0,82925 | -0,428929 |
| 010 | 0,364998 | -0,82925 | -0,428920 |
| 110 | 0,364998 | -0,82925 | -0,428920 |
| 101 | -0,63384 | 0,15502 | 0,847544 |
| 011 | -0,63384 | 0,15502 | 0,847544. |

**Table 2.** Magnetic anisotropy of Gadolinium using GGA approximation

Table 2 indicates that for bulk Gadolinium, the preferred magnetization axis is along the c-axis. In contrast, when considering the thin Gadolinium layer, the energy is minimized along the [100], [010], and [110] axes, signifying that the a-axis and b-axis serve as the favored magnetization directions in Gadolinium thin films. This finding suggests that electrons in our thin layer of Gadolinium can move easily along the (ab) plane. This insight guided our decision regarding the orientation of the 100nm-thick Gadolinium layer in our experimental study.

## IV. Conclusion

In this paper, we studied the structural and magnetic properties of both bulk and thin film forms of Gadolinium in order to show the main characteristics sought for magnetic refrigeration. The key problem that we have addressed after the simulation of these materials is the potential of thin film materials for cooling electronic circuits. Each atom of a material has a net magnetic moment which is the sum of the spin of its electrons. The sum of the net magnetic moment of all the atoms contained in a unit of volume of the magnetic material represents magnetization. This research provides valuable insights into the possible applications and broader implications of Gadolinium within the domain of magnetic refrigeration technology.